\documentclass[12 pt, amsfonts, amssymb,color]{article}

\usepackage{amsmath,amssymb,amsfonts,latexsym,float,graphics,epsfig}

\begin{document}
\renewcommand\theequation{\arabic{section}.\arabic{equation}}
\catcode`@=11 \@addtoreset{equation}{section}
\newtheorem{axiom}{Definition}[section]
\newtheorem{theorem}{Theorem}[section]
\newtheorem{axiom2}{Example}[section]
\newtheorem{lem}{Lemma}[section]
\newtheorem{prop}{Proposition}[section]
\newtheorem{cor}{Corollary}[section]
\newcommand{\be}{\begin{equation}}
\newcommand{\ee}{\end{equation}}
\newcommand{\equal}{\!\!\!&=&\!\!\!}
\newcommand{\rd}{\partial}
\newcommand{\g}{\hat {\cal G}}
\newcommand{\bo}{\bigodot}
\newcommand{\res}{\mathop{\mbox{\rm res}}}
\newcommand{\diag}{\mathop{\mbox{\rm diag}}}
\newcommand{\Tr}{\mathop{\mbox{\rm Tr}}}
\newcommand{\const}{\mbox{\rm const.}\;}
\newcommand{\cA}{{\cal A}}
\newcommand{\bA}{{\bf A}}
\newcommand{\Abar}{{\bar{A}}}
\newcommand{\cAbar}{{\bar{\cA}}}
\newcommand{\bAbar}{{\bar{\bA}}}
\newcommand{\cB}{{\cal B}}
\newcommand{\bB}{{\bf B}}
\newcommand{\Bbar}{{\bar{B}}}
\newcommand{\cBbar}{{\bar{\cB}}}
\newcommand{\bBbar}{{\bar{\bB}}}
\newcommand{\bC}{{\bf C}}
\newcommand{\cbar}{{\bar{c}}}
\newcommand{\Cbar}{{\bar{C}}}
\newcommand{\Hbar}{{\bar{H}}}
\newcommand{\cL}{{\cal L}}
\newcommand{\bL}{{\bf L}}
\newcommand{\Lbar}{{\bar{L}}}
\newcommand{\cLbar}{{\bar{\cL}}}
\newcommand{\bLbar}{{\bar{\bL}}}
\newcommand{\cM}{{\cal M}}
\newcommand{\bM}{{\bf M}}
\newcommand{\Mbar}{{\bar{M}}}
\newcommand{\cMbar}{{\bar{\cM}}}
\newcommand{\bMbar}{{\bar{\bM}}}
\newcommand{\cP}{{\cal P}}
\newcommand{\cQ}{{\cal Q}}
\newcommand{\bU}{{\bf U}}
\newcommand{\bR}{{\bf R}}
\newcommand{\cW}{{\cal W}}
\newcommand{\bW}{{\bf W}}
\newcommand{\bZ}{{\bf Z}}
\newcommand{\Wbar}{{\bar{W}}}
\newcommand{\Xbar}{{\bar{X}}}
\newcommand{\cWbar}{{\bar{\cW}}}
\newcommand{\bWbar}{{\bar{\bW}}}
\newcommand{\abar}{{\bar{a}}}
\newcommand{\nbar}{{\bar{n}}}
\newcommand{\pbar}{{\bar{p}}}
\newcommand{\tbar}{{\bar{t}}}
\newcommand{\ubar}{{\bar{u}}}
\newcommand{\utilde}{\tilde{u}}
\newcommand{\vbar}{{\bar{v}}}
\newcommand{\wbar}{{\bar{w}}}
\newcommand{\phibar}{{\bar{\phi}}}
\newcommand{\Psibar}{{\bar{\Psi}}}
\newcommand{\bLambda}{{\bf \Lambda}}
\newcommand{\bDelta}{{\bf \Delta}}
\newcommand{\p}{\partial}
\newcommand{\om}{{\Omega \cal G}}
\newcommand{\ID}{{\mathbb{D}}}
\newcommand{\pr}{{\prime}}
\newcommand{\prr}{{\prime\prime}}
\newcommand{\prrr}{{\prime\prime\prime}}
\newcommand{\bel}{\begin{equation}\label}
\title{On a geometric description of  time dependent singular Lagrangians with applications to biological systems}
\author{Sudip Garai\footnote{E-mail sudip.dhwu@gmail.com} and A Ghose-Choudhury\footnote{E-mail aghosechoudhury@gmail.com}\\
Department of Physics \\
Diamond Harbour Women's University,\\ D. H Road, Sarisha,
West-Bengal 743368, India\\
\and
Partha Guha\footnote{E-mail: partha.guha@ku.ac.ae}\\
Department of Mathematics \\
Khalifa University\\ P.O. Box 127788, Abu Dhabi, UAE \\}
\date{}
\maketitle
\smallskip
\smallskip
\begin{abstract}
\textit{We consider certain  analytical features of   a stochastic model that can explain among other things competition among species and simultaneous predation on the competing species from a geometric perspective which allows for a systematic description of models admitting singular Lagrangians. The model equations are shown to admit a Jacobi Last Multiplier which in turn allows for the construction of a Lagrangian. The Lagrangian is of singular nature so that construction of the Hamiltonian via a Legendre transformation is not possible. A Hamiltonian description of the model therefore requires the introduction of Dirac brackets. Explicit results are presented for the "Kill the winner" model and its reductions.}

\end{abstract}
\section{Introduction}
 Population dynamics is a well established field and can lay claim to having increased our understanding of several phenomena is diverse areas. However, most theoretical models of population dynamics do not offer a  satisfactory explanation for  the so-called diversity paradox in nature. Roughly speaking when several species  compete for the same finite resource, a theory called competitive exclusion indicates that one species will outperform the others and drive them to extinction, thereby limiting biodiversity. However, nature does not quite evolve  in this manner and shows up a greater degree of diversity  then predicted.
  Goldenfeld and Xue developed a stochastic model \cite{KTW} that accounts for multiple factors observed in ecosystems, including competition among species and simultaneous predation on the competing species. Using bacteria and their host-specific viruses as an example, they showed that as the bacteria evolve defenses against the virus, the virus population also evolves to combat the bacteria. This "arms race" leads to a diverse population of both and to boom-bust cycles when a particular species dominates the ecosystem then collapses- the so-called "Kill the Winner" (KtW) phenomenon. This coevolutionary arms race is sufficient to yield a possible solution to the diversity paradox.\\
   It is of interest to look at such systems from a purely dynamic point of view as they provide interesting concrete examples of first-order differential systems which have physical relevance \cite{TF, Go, Paine}.
Of late there has been a lot of  interest in the study of  systems with singular Lagrangians which have arisen in
	the field of quantitative biology. Several well known models such as the Lotka-Volterra prey-predator system, the host-parasite model, the Bailey model etc. are the examples of  systems that are described by singular Lagrangians, i.e. by  Lagrangians
that do not depend quadratically on the velocity parameter \cite{NT}. A major problem in describing the dynamics of such systems on
the phase space is in defining the appropriate canonical momentum. In order to analyze the dynamics of such systems, Dirac formulated a method which basically makes the use of a constrained surface on which the Poisson bracket may be suitably defined. Roughly speaking this restricted Poisson bracket defined on a sub-manifold of the complete phase space of the system is usually referred to the Dirac bracket.\\

 A Lagrangian is said to be singular if the corresponding Hessian matrix is singular. The singular nature of the Hessian matrix prevents us from solving for the velocities in terms of the conjugate momenta and the coordinates. This in turn prevents us from constructing the Hamiltonian by means of a Legendre transformation.
In order to deal with singular Lagrangians, and motivated by the need to study the quantization of electromagnetism,
	P.A.M. Dirac developed a constraint algorithm, known as the Dirac-Bergmann algorithm, that allows us to construct the dynamics of the system \cite{Dirac}. This constraint algorithm has been later geometrized by M.J. Gotay and J.M. Nester \cite{GN}.

Dirac introduced two kind of constraints namely, that of the first and second class.  The second class of constraints allows us to define a Poisson bracket (called Dirac bracket) that gives the dynamics of the con-strained system just as in the classical case with the canonical Poisson bracket.

\bigskip

It is evident that if the Lagrangian is linear in the velocities then it is singular in character because the Hessian matrix is trivial. As already mentioned a particularly rich field in which the dynamics may be described by singular Lagrangians arises in models describing ecological systems, predator-prey models etc. Generically speaking all these Lagrangians are time dependent. In dealing with such time dependent systems the methods of cosymplectic geometry are found to be  particularly useful \cite{esenguha,deLeon1,deLeon2}, and will be applied here to the case of  time-dependent singular Lagrangian systems.

\medskip

In this article we focus on an interesting stochastic model dealing with the problem of biological diversity. The so called "Kill the Winner" (KtW) hypothesis has been quite extensively studied by many authors. The present analysis is based on the model equations introduced in \cite{KTW}. While most studies of such systems have relied on a numerical approach we have adopted a more analytical view expanding on our previous work with biological systems. Several well known systems such as the Lotka-Volterra, Host-parasite, Kermack-McKendrick models may be shown to be special cases of the general KtW model studied here. It is interesting to note that the stochastic model equations admit a Jacobi Last Multiplier (JLM) which in turn enables us to set up a Lagrangian description of the system. However, it is found that the Lagrangian obtained is singular in character so that a proper Hamiltonian description of the system in phase space requires the introduction of Dirac brackets.\\
The paper is organized as follows: In section 2 we outline the procedure for deriving a Lagrangian linear in the velocities (and hence singular) assuming the existence of a JLM. A geometric description of such singular Lagrangians is then presented followed by an introduction to the notion of Dirac brackets.  In section 3 we introduce the "Kill the Winner" (KtW) model and considering its simplest version derive the explicit expression for the JLM. This is done to motivate the subsequent general results. The Lagrangian  for the general KtW model is then derived and the ingredients required for constructing the Dirac brackets is  presented. This is followed by an explicit calculation of the Dirac brackets for some specific cases  such as the Lotka-Volterra model with and without competition and the Gierer-Meinhardt model of pattern formation.

\section{Preliminaries}

The time evolution associated with a time-dependent mechanical system is
usually represented by the flow of a vector field defined in ${\Bbb R} \times TQ$, where
$Q$ is a smooth differentiable manifold which represents the original configuration
space of the system. This is a space of $1$-jets
of the trivial bundle $\pi : {\Bbb R} \times Q \to {\Bbb R}$, i.e., $J^1\pi = {\Bbb R} \times TQ$,
which in terms of local coordinates given by $(t,q^i,v^i) \in J^1\pi$ \cite{Saunders}.
Let $L : TQ \times {\Bbb R} \to {\Bbb R}$ be a Lagrangian function, it is regular when the Hessian matrix
$\big( \frac{\partial ^2L}{\partial \dot{q}^i\partial \dot{q}^j} \big)$ is non-singular.

\smallskip

The local $1$-forms, $\theta^i = dq^i - v^idt$,
$i=1, \cdots , n$, constitute a local basis for the contact $1$-forms.
For the case of $n = 2$, given the system of equations
\be\label{a.1}\frac{dx}{dt}=f(x, y, t), \qquad \frac{dy}{dt}=g(x, y, t),\ee
the wedge product of the Cartan one-forms, $\theta
^{x}=dx-f(x,y,t)dt\;$ and $\;\theta^{y}=dy-g(x,y,t)dt$ is defined as
\begin{equation}
\omega=\theta^{x}\wedge\theta^{y}. \label{alpha}%
\end{equation}

\smallskip

\bigskip

The Lagrangian (or Hamiltonian) formulation in the extended phase space is characterized by the Poincar\'{e}-Cartan one-form.
We briefly introduce the basic definition of a vertical endomorphism which is used
to define the Poincar\'e-Cartan form \cite{Car1, Car2}.
Let $ v = v^i\frac{\partial}{\partial q^i} \in T_qQ$
and $V = \dot{q}^i \frac{\partial}{\partial q^i} + \dot{v}^i\frac{\partial}{\partial v^i} \in T_v(TQ)$
such that $\tau_{TQ}(V) = v$.
In local coordinates, $\tau_{TQ}(q^i,v^i,\dot{q}^i, \dot{v}^i ) = (q^i,v^i)$.
If $\tau_Q: v \in TQ \mapsto q \in Q$ denotes the natural projection, then, given a tangent vector
$V \in T_v(TQ)$,
we have $\tau_{TQ}(V) = v$. Let $v \in T_qQ$ be a vector tangent to $Q$ at some point $q \in Q$. Then
the vertical lift of $v$ at a {\it point} $w \in T_qQ$ is the tangent vector $v_{w}^{V} \in T_w(TQ)$ given by
$$u_{w}^{V} (g) = \frac{d}{dt} g(w + tv)|_{t=0} \qquad \forall g \in C^{\infty}(T_qQ).$$ In local coordinates
of $T^2Q$ (or $T(TQ)$) this is given by $ v_{w}^{V} = v^i\frac{\partial}{\partial v^i}|_w$, which is the Liouville vector
field over $TQ$.
The vertical endomorphism is the linear map $S : T^2Q \to T^2Q$ for which any vector $V \in T^2Q$ yields
$S(V) = ((T_u\tau_Q)(V))^V$, where $v = \tau_{TQ}(V) \in TQ$. From $L$, and using the local expression of $S = dq^i \otimes
\frac{\partial}{\partial q^i}$, we can define the Poincar\'e-Cartan 1-form $\Theta_L = dL \circ S$.

\medskip

There is a natural extension of vertical endomorphism to the non-autonomous case. In fibred coordinates,
$(t,q^i,v^i)\in {\Bbb R} \times TQ$, $S$ is given as $S = (dq^i - v^idt) \otimes \partial/ \partial v^i$ \cite{Crampin}.
The most important
objects associated with the time-dependent Lagrangian are the Poincar\'e-Cartan one and two forms defined by
\be \Theta_L = dL \circ S + L dt, \qquad \Omega_L := - d\Theta_L. \ee
Their local expressions are
\be \Theta_L = \frac{\partial L}{\partial v^i}(dq^i - v^i dt) + Ldt =
\big(L - v^i\frac{\partial L}{\partial v^i}\big) dt
+  \frac{\partial L}{\partial v^i}dq^i \equiv -E_Ldt + d_JL,
\ee
where $E_L$ is the energy function associated to $L$ and $d_J$ is the operator associated to almost tangent structure on $TQ$.

Then the Poincar\'e-Cartan $2$-form associated to $L$ is given by
\be\label{PC2} \Omega_L =  -d\big( \frac{\partial L}{\partial v^i}\big) \wedge dq^i +
d\big( v^i\frac{\partial L}{\partial v^i} - L\big) \wedge dt, \quad \hbox{ since } \,\,\,\, \Omega_L := - d\Theta_L.\ee

\bigskip

{\bf Cosymplectic condition and the Jacobi last multiplier :}\,  A cosymplectic manifold is a triple $({\mathcal{M}},\eta, \omega)$
consisting of a smooth $(2n+1)-$dimensional manifold ${\mathcal{M}}$ with a closed $1$%
-form $\eta$ and a closed $2$-form $\Omega$ such that $\eta \wedge \Omega^n $ is a non-vanishing volume form on ${\mathcal{M}}$.
There is a distinguished vector field $\xi$, called a Reeb field, on $({\mathcal{M}},\eta, \Omega)$ which is uniquely
determined by the conditions
\begin{equation} \label{xi}
i_\xi\eta = 1, \qquad i_{\xi}\Omega = 0.
\end{equation}

We consider the cosymplectic structure $(\Omega_L,dt)$ on $TQ \times {\Bbb R}$,
where the Reeb vector $\xi = \frac{\partial}{\partial t}$.

\medskip

 From (\ref{PC2}) we obtain
$$
\Omega_L = -\big( \frac{\partial^2L}{\partial v^i \partial t} + v^j \frac{\partial^2L}{\partial v^j \partial q^i}
- \frac{\partial L}{\partial q^i} \big) dt \wedge dq^i + v^j \frac{\partial^2L}{\partial v^i \partial v^j} dv^i \wedge dt
$$
$$
- \frac{\partial^2L}{\partial v^i \partial q^j} dq^j \wedge dq^i - \frac{\partial^2L}{\partial v^i \partial v^j}dv^i \wedge dq^j.
$$

Let us restrict ourselves to $1$-dimension, so that $ i=j=1$, then we obtain
\be
\Omega_{L}^{1} =  -\big( \frac{\partial^2L}{\partial v \partial t} + v \frac{\partial^2L}{\partial v \partial q}
- \frac{\partial L}{\partial q} \big)dt \wedge dq +  v \frac{\partial^2L}{\partial v^2} dv \wedge dt
- \frac{\partial^2L}{\partial v^2}dv \wedge dq.
\ee
Contraction  with the vector field $\xi = \partial/\partial t$ then yields
$$
i_{ \partial/\partial t}\Omega_{L}^{1} =  - \Big(\frac{\partial^2L}{\partial v^2}\Big) \big( v dv - Fdq)
= \big( \frac{d}{dt}(\frac{\partial L}{\partial v}) - \frac{\partial L}{\partial q}) \big) dq.
$$
Thus $i_{ \partial/\partial t}\Omega_{L}^{1} = 0$ leads to the
the existence of the Jacobi last multiplier \cite{Whi} or the celebrated Helmholtz criteria in a
differential form and the JLM $\mu$  is given by
$$ \mu =  \big(\frac{\partial^2L}{\partial v^2}\big). $$

\subsection{Hamiltonian formulation and JLM}

The Hamiltonian formulation in the extended phase space is characterized by the Poincar\'{e}-Cartan one-form $\Theta=PdQ-Hdt$
from which we may define a closed two-form $\Omega$ by minus the exterior
derivative of $\Theta$, that is,
\begin{equation}
\Omega=dQ\wedge dP+dH\wedge dt. \label{5.8}%
\end{equation}
It follows that, the two-form $\omega$ in Eq.(\ref{alpha})\ is proportional to
$\Omega$ with
\begin{equation}
M\omega=\Omega\label{alpha-omega}%
\end{equation}
for some function $M$, called the Jacobi Last Multiplier \cite{Whi}. Since $\Omega$ is
closed, $M\omega$ must be closed. This leads to the following differential
equation
\begin{equation}
\partial_{t}M+\partial_{x}(Mf)+\partial_{y}(Mg)=0 \label{5.10}%
\end{equation}
determining $M$. After solving equation (\ref{5.10}), the canonical
coordinates can be obtained by substituting $M$ into Eq.(\ref{alpha-omega}).
To determine the Hamiltonian function $H$, there are two possible cases
depending whether $\partial_{t}M$ vanishes or not.

If the multiplier $M$ does not depend on the time $t$ then the first term on
the left hand side of Eq.(\ref{5.10}) drops. In this case, from the equality
$M\omega=\Omega$, we arrive at%
\begin{equation}
M\left(  fdy-gdx\right)  =dH, \label{5.11}%
\end{equation}
which relates $M$ with the Hamiltonian function $H$.

When the multiplier $M$ depends on time explicitly, we introduce two auxiliary
functions $\phi$ and $\psi$ such that the structure of Eq.(\ref{5.11}) is
preserved, that is, for a pair of functions $\psi$ and $\phi$ we have
\begin{equation}
M((f-\psi)dy-(g-\phi)dx)=dH+\vartheta dt, \label{5.13}%
\end{equation}
for a real valued function $\vartheta$ of $(Q,P,t)$. This occurs
if the condition
\begin{equation}
\partial_{x}(M(f-\psi))+\partial_{y}(M(g-\phi))=0 \label{5.12}%
\end{equation}
is satisfied. By adding and subtracting two-form $M\psi dy\wedge dt-M\phi
dx\wedge dt$ into Eq.(\ref{alpha-omega}), we obtain%
\begin{align*}
M\omega &  =M\theta^{x}\wedge\theta^{y}=Mdx\wedge dy+M(fdy-gdx)\wedge dt\\
&  =Mdx\wedge dy+M(fdy-gdx)\wedge dt\pm\left(  M\psi dy\wedge dt-M\phi
dx\wedge dt\right) \\
&  =M(dx-\psi dt)\wedge(dy-\phi dt)+M\left(  (f-\psi)dy-(g-\phi)dx\right)
\wedge dt,
\end{align*}
where the first two-form in the last line is the symplectic two-form
\begin{equation}
M(dx-\psi dt)\wedge(dy-\phi dt)=dQ\wedge dP, \label{Canoor}%
\end{equation}
and the latter can be obtained by taking the exterior product of both sides of
Eq.(\ref{5.13}) by $dt.$ Note that, substitutions of the auxiliary functions
$\psi$ and $\phi$, and the multiplier $M$ into Eq.(\ref{5.13}) will enable us
to find the Hamiltonian function whereas substitutions into Eq.(\ref{Canoor})
will determine the canonical coordinates.

\subsection{Dirac bracket}

Let $L$ be now a singular Lagrangian function, in this case $\Omega_L$ is no longer symplectic. By performing the constraint
algorithm of Dirac-Bergmann-Gotay-Nestor we obtain a sequence of constraint submanifolds having nested structure.
Suppose $L$ is almost regular, i.e., the Legendre map $FL : TQ \to M_1$
is a surjective submersion map satisfying $M_1 = FL(TQ) \subseteq T^{\ast}Q$.
The first constraint submanifold $M_1$ is endowed with a Hamiltonian $H_1 : M_1 \to {\Bbb R}$.
Let $H$ be an extension of $H_1$, then the constrained Hamiltonian defined on $T^{\ast}Q \times {\Bbb R}$ is given by
\be H_T = H + u_a x_a, \ee
where $x^a$ are a set of constraints defining $M_1$ and $u_a$ are Lagrange multipliers.
It is customary to demand that the primary constraints should be preserved along the evolution of the system,
geometrically this means $X_{H_T}$ is tangent to $M_1$, i.e.
$$
X_{H_T}(x_a) = \{H,x_a \} + u_a \{x_b,x_a \}|_{M_1} = 0.
$$

\smallskip

Let $M_c$ be the final constraint submanifold by imposing $2m$ independent constraints, this yields $2(n-m)$-dimensional symplectic manifold
$M_c \subset M$. In the neighbourhood of a point $p \in M$, choose coordinates $x_1, \cdots x_{2n} \in M$, such that
$M_c$ is given by $x_1 = 0, \cdots, x_{2m-1} = 0, x_{2m} = 0,$ constraints are forcing the system to lie on $M_c$.
Instead of functions, we can supply $2(n-m)$ vector fields whose kernel is $M_c$.
Thus $x_{2m+1}, \cdots ,  x_{2n}$ provide local coordinates on $M_c$.

\smallskip

Let us define the matrix $$ C^{rs}(x) = \{x_r,x_s\}, \qquad r,s = 1, \cdots 2m, $$ and suppose $f,g \in C^{\infty}(TQ \times {\Bbb R})$
be the smooth
functions, and let $f^{\prime}, g^{\prime}$ be their restriction to $M_c$.

\begin{axiom}
	A function $f \in C^{\infty}(TQ \times {\Bbb R})$ is called first class if $\{f,x_r\}|_{M_c} = 0, \forall r$,
	and this set of first class  constraints form a subalgebra with respect to Poisson bracket.
All other constraints are known as second class constraints,  and satisfy
$\{x_r,x_s\} = C_{rs}$.
Then the Dirac bracket is defined by
\be
\{f^{\prime}, g^{\prime} \}_{D} := \{f,g\} - \sum_{r,s =1}^{2m}\{f,x_s\}[C_{rs}]^{-1}\{x_s,g\},
\ee
where the double sum is taken for all second class constraints. This bracket satisfies Jacobi identity.

\end{axiom}

The vector field associated to the Dirac bracket is given as
\be
X_D = X_f - \{f,x_r \}C^{rs}X_{x_s}.
\ee

 For any second class constraint
$$
\{f^{\prime}, x_k \}_{D} := \{f,x_k\} - \sum_{r,s =1}^{2m}\{f,x_r\}[C_{rs}]^{-1}\{x_s,x_k\}\{f,x_s\}
$$
$$ = \{f,x_k\} - \sum_{r,s =1}^{2m}\{f,x_r\}[C_{rs}]^{-1}[C_{sk}] $$
$$= \{f,x_k\} - \sum_{r=1}^{2m}\{f,x_r\}\delta_{rk} = \{f,x_k\} - \{f,x_k\} =0.
$$
which shows that the flow generated by the second class constraint vanishes, hence it does not leave the constraint manifold.
Thus Dirac bracket provides a modification of the Poisson bracket so as to ensure that the Hamiltonian flow generated by
the constraints with respect to the new Poisson structure is tangent to the constraint manifold \cite{Ibort}.

\smallskip

Using bivector formalism, the Dirac bivector is given as
\be
\Pi_D := \Pi + \frac{1}{2} [C_{rs}]^{-1} {\cal X}_r \wedge {\cal X}_s,
\ee
where $\Pi$ stands for Poisson bivector and
$$ {\cal X}_r = \{\cdot , x_r \} = - (\partial_i x_r) \Pi^{ij}\partial_j. $$

\smallskip

Let $\{p_j, q^j \}$, $1 \leq j \leq 2n$, denote a set of dynamical variables, $\{u^a\}$,
$1 \leq a \leq 2m$, set of Lagrange multipliers, and $\{x_a(p,q)\}$ a set of constraints.
The total Hamiltonian is given by
\be
H_{Tot} = H(p,q) + u_ax_a(p,q).
\ee
Then the dynamics of a constrained system can be obtained from the action principle
$$
S = \int [p_j\dot{q}^j - H(p,q) - u_ax_a(p,q)] dt.
$$
The resultant equations that arise from the action read
\be
\dot{p}_i = - \frac{\partial H}{\partial q^i} + u_a \frac{\partial x_a}{\partial q^i}, \,\,\,\,\,\,
\dot{q}^i =  \frac{\partial H}{\partial p_i} - u_a \frac{\partial x_a}{\partial p_i}, \,\,\,\,\,\, \dot{x_a} = 0.
\ee
The (constrained) Hamiltonian equation for any arbitrary smooth function $g(p,q) \in C^{\infty}(M)$ is given
by
\be
\dot{g} = \{ g, H(p,q) + u_ax_a \} = \{ g, H(p,q)\} + \{ g, u_ax_a \} \approx \{ g, H(p,q)\} + u_a\{ g, x_a \},
\ee
where the $\approx $ symbol means that we should not substitute $x_a = 0$ before evaluating Poisson bracket.
This is called the weak equation. For a variation of the weak equation $\dot{g} \approx \{ g, H(p,q) + u_ax_a \}$,
the variation of LHS is not equal to the variation of RHS.

\smallskip

Let us assume that all the constraints are second class, then the corresponding undetermined coefficients
are given by
\be\label{coeff} u_b = - \{H,x_a\}[C_{ab}]^{-1}. \ee
We can express the equation of motion with second class constraints in terms of Dirac bracket.

\begin{prop}
The equation of motion for a system with second class constraints can be expressed by
\be
\dot{g} \approx \{g, H\}_{D} = \{g,H\} - \sum_{a,b} \{g,x_a\}[C_{ab}]^{-1}\{x_b, H\}.
\ee
\end{prop}
This can be proved easily using (\ref{coeff}).

\section{Singular Lagrangians of biological systems using Jacobi last multiplier and Dirac bracket}

Let us briefly recall the procedure described in \cite{NT, CGK, CG} for finding Lagrangians for a planar system of ODEs from
a knowledge of the last multiplier. We assume that the  system
$$ \frac{dx}{dt}=f(t, x, y), \qquad  \frac{dy}{dt}=g(t, x, y) $$
 admits a Lagrangian which is linear
in the velocities, so that
 \be\label{a.3}L(t, x,y,\dot{x},
 \dot{y})=F(t,x,y)\dot{x}+G(t,x,y)\dot{y}-V(t,x,y).\ee
Then the Euler-Lagrange equations of motion yields
\be\label{a.4}\dot{y}=\left(\frac{F_t+V_x}{G_x-F_y}\right)=g(t,x,y), \qquad \dot{x}=-\left(\frac{G_t+V_y}{G_x-F_y}\right)=f(t,x,y).\ee
Here the subscripts on $F,G$ and $V$ denote the partial derivatives
while the over-dots represent the derivative with respect to time. It is obvious that one must have $G_x\ne F_y$.
In order to introduce the notion of Jacobi's last multiplier we assume that
$G_x=-F_y$ and  assign a common value,
\be\label{a.6}\mu(t,x,y):=G_x=-F_y.\ee From (\ref{a.4}) we have
\be 2\mu f(t,x,y)  =-(G_t+V_y), \qquad 2\mu
g(t,x,y) =(F_t+V_x).\ee

It is clear that the construction
$$\frac{\partial}{\partial x}\left(2\mu f\right)+\frac{\partial}{\partial y}\left(2\mu
  g\right)$$ leads to the Jacobi last multiplier equation \cite{Whi}.
Thus we
see that given the solution of this equation one can easily construct
from (2.4) the coefficient functions $F$ and $G$ occurring in the
expression for the Lagrangian since \be\label{a.8}F(t,x,y)=-\int
\mu(t,x,y) dy\;\;\;\mbox{and}\;\;\;G(t,x,y)=\int \mu(t,x,y)dx.\ee Once
these functions are determined one can obtain an expression for the
partial derivatives of $V$ from (\ref{a.4}) as follows
\be
\frac{\partial V}{\partial x}  =2\mu(t,x,y)
g(t,x,y)+\frac{\partial}{\partial t}\left(\int \mu
dy\right), \quad \frac{\partial V}{\partial y}
=-2\mu(t,x,y) f(t,x,y)-\frac{\partial}{\partial t}\left(\int \mu
dx\right).\ee  It is easy to check
the equality of  the mixed derivatives.

\subsection{``Kill the Winner'' Model}

The ``Kill the winner (KtW)'' hypothesis  attempts to develop a stochastic model for the problem of biological diversity  in Nature where the host-specific predators control the prey population of each species by preventing a winner from emerging. This natural phenomenon maintains the equilibrium coexistence of all the existing species in the system. An  individual-level stochastic model in which predator-prey coevolution promotes the high diversity of
the ecosystem by generating a persistent population flux of species has been developed in \cite{KTW}. For a single species  the model is described by the following system of differential equations \cite{KTW}:
\be\label{ktw1a}
\dot{x} = a_1 x - b_1 x^2 - c_1 xy\ee
\be\label{ktw1b}\dot{y} = a_2 xy - b_2 y
\ee
where, $a_1$, $a_2$, $b_1$, $b_2$, $c_1$ are the constants and $x(y)$ represents the bacterial(viral strains) density of the system. The Jacobi last multiplier for this system can be written as $M=e^{\gamma t}x^{\alpha}y^{\beta}$ with
\begin{eqnarray}
\alpha &=&-1\\
\beta &=& \frac{b_1 - a_2}{a_2}\\
\gamma &=& \frac{b_1 b_2}{a_2}
\end{eqnarray}
The generalized KTW model is given by the system of equations
\begin{eqnarray}
\dot{x}_i &=& b_i x_i - \sum_{j=1}^m e_{ij}x_ix_j - p_ix_iy_i\\
\dot{y}_i &=& q_i x_i y_i-d_i y_i,\;\;\;i=1,...,m
\end{eqnarray}
The Jacobi last multiplier for the system is given by
\be
M=\prod_{i=1} ^{m} M_i
\ee
with $M_i = e^{\gamma_i t} y_i ^{\sigma_i}/x_i$. Here $\gamma_i = e_{ii}d_i/q_i$ and $\sigma_i = -1 + e_{ii}/q_i$. Assuming the singular Lagrangian to be
\be
L=\sum_{k=1} ^{m} \left[F_k (x,y) \dot{x_k} + G_k (x,y) \dot{y}\right] - U(x,y)
\ee
we have on substituting this into the Euler-Lagrange equation:
\be
\sum_{k=1} ^{m} \left[\left(F_{i x_{k}} - F_{k x_{i}}\right) \dot{x} + \left(F_{i y_{k}} - G_{k x_{i}}\right) \dot{y} \right] = - \left(\frac{\partial U}{\partial x_i} + \frac{\partial F_i}{\partial t}\right),
\ee
\be
\sum_{k=1} ^{m} \left[\left(G_{i x_{k}} - F_{k y_{i}}\right) \dot{x} + \left(G_{i y_{k}} - G_{k y_{i}}\right) \dot{y} \right] = - \left(\frac{\partial U}{\partial y_i} + \frac{\partial G_i}{\partial t}\right).
\ee
The above equations can be written in  matrix form as
$$\left(\begin{array}{cc}
\dot{X}\\ \dot{Y}\end{array}\right)=-\left(\begin{array}{cccc} \mathbb{A} & \mathbb{B}\\
-\mathbb{B}^T & \mathbb{D}\end{array}\right)^{-1}\left(\begin{array}{cc}
P\\ Q\end{array}\right) = -\frac{1}{\Delta}\left(\begin{array}{cccc} \mathbb{D} & -\mathbb{A}^{-1}\mathbb{B}\mathbb{A}\\
\mathbb{D}^{-1}\mathbb{B}^T\mathbb{D} & \mathbb{A}\end{array}\right)\left(\begin{array}{cc}
P\\ Q\end{array}\right)$$
where
$$X=\left(\begin{array}{cccc} x_1\\ \vdots\\x_m\end{array}\right), \;\;\;Y=\left(\begin{array}{cccc} y_1\\ \vdots\\y_m\end{array}\right), \;\;\;P=\left(\begin{array}{cccc} \frac{\partial F_1}{\partial t}+\frac{\partial U}{\partial x_1} \\ \vdots\\\frac{\partial F_m}{\partial t}+\frac{\partial U}{\partial x_m}\end{array}\right), \;\;\;Q=\left(\begin{array}{cccc} \frac{\partial G_1}{\partial t}+\frac{\partial U}{\partial y_1}\\ \vdots\\\frac{\partial G_m}{\partial t}+\frac{\partial U}{\partial y_m}\end{array}\right)$$
Now we make the assumption that $\mathbb{A} = \mathbb{D} = 0$ and $\mathbb{B}$ is diagonal with $\mathbb{B}_{ii} = F_{i y_i} - G_{i x_i}$. Therefore we can write
\be
\dot{y}_i = - \frac{\left(\frac{\partial U}{\partial x_i} + \frac{\partial F_i}{\partial t}\right)}{\left(\frac{\partial F_i}{\partial y_i} - \frac{\partial G_i}{\partial x_i}\right)}
\ee
\be
\dot{x}_i = \frac{\left(\frac{\partial U}{\partial y_i} + \frac{\partial G_i}{\partial t}\right)}{\left(\frac{\partial F_i}{\partial y_i} - \frac{\partial G_i}{\partial x_i}\right)}
\ee
Next let us  assume,
\be
\frac{\partial F_i}{\partial y_i} =  - \frac{\partial G_i}{\partial x_i} = M_i
\ee
therefore,
\be
\dot{y}_i = - \frac{1}{2M_i} \left(\frac{\partial U}{\partial x_i} + \frac{\partial F_i}{\partial t}\right)
\ee
\&
\be
\dot{x}_i = - \frac{1}{2M_i} \left(\frac{\partial U}{\partial y_i} + \frac{\partial G_i}{\partial t}\right)
\ee
It follows that
\be
F_i = \int M_i d y_i + f_i (t,x_i) = \frac{e^{\gamma_i t}}{x_i} \frac{y_i ^{\sigma_i +1}}{\left(\sigma_i +1\right)} + f_i (t,x_i)
\ee
and
\be
G_i = - \int M_i d x_i + g_i (t,y_i) =  - e^{\gamma_i t} y_i ^{\sigma_i} \log x_i + g_i (t,y_i)
\ee
Now from the above relations we can see that
\be
\frac{\partial U}{\partial x_i} = - \frac{e^{\gamma_i t} y_i ^{\sigma_i +1}}{x_i} \left[2\left(q_i x_i - d_i \right) + \frac{\gamma_i}{\left(\sigma_i +1\right)}\right]
\ee
and
\be
\frac{\partial U}{\partial y_i} =  e^{\gamma_i t} y_i ^{\sigma_i} \left[2 b_i - 2 \sum_{j=1} ^{m} e_{ij} x_j - 2 p_i y_i + \gamma_i \log x_i\right]
\ee
The solution for the potential $U$ can be obtained from the above set of equations as
\be
U = - \sum_{i=1} ^{m} \left[e^{\gamma_i t} y_i ^{\sigma_i +1} \left\{2 q_i x_i - d_i \log x_i + \frac{2 p_i y_i}{\left(\sigma_i +2\right)}\right\} + e^{\gamma_i t} y_i ^{\sigma_i +1} \frac{2 q_i}{e_{ii}} \left(\sum_{\substack{j=1,\\ j \neq i}} ^{m} e_{ij} x_j - b_i\right)\right]
\ee
Hence the Lagrangian for the system can be written as
\begin{eqnarray}
L = \sum_{k=1} ^{m} e^{\gamma_k t} y_k ^{\frac{e_{kk}}{q_k}} \left[\frac{q_k}{e_{kk}} \frac{\dot{x}_k}{x_k} - \log x_k \frac{\dot{y}_k}{y_k} + \left\{2 q_k x_k - d_k \log x_k + \frac{2 q_k}{e_{kk}} \left(\sum_{\substack{j=1, \\ j \neq k}} ^{m} e_{kj} x_j - b_k\right) \right. \right. \nonumber\\
\left. \left. + 2 p_k y_k \left(\frac{q_k}{e_{kk}} +1 \right)\right\}\right]
\end{eqnarray}

The expression for the Hamiltonian turns out to be
$$H=\sum_{k=1}^m\left[\frac{\partial L}{\partial \dot{x}_k}\dot{x}_k +\frac{\partial L}{\partial \dot{y}_k}\dot{y}_k\right] -L=U(x,y)$$
$$=-\sum_{k=1}^me^{\gamma_k t} y_k ^{\frac{e_{kk}}{q_k}} \left[2q_kx_k-d_k\log y_k+\frac{2q_k}{e_{kk}}\left(\sum_{\substack{j=1, \\ j \neq k}} ^{m} e_{kj} x_j - b_k\right)+2p_ky_k\left(\frac{q_k}{e_{kk}}+1\right)\right]$$
and is independent of the velocities as is natural for such singular Lagrangians.

Let us now identify the primary constraints of the system:
It is easily seen that the conjugate momenta are given by
$$p_k^x=\frac{\partial L}{\partial \dot{x}_k}=F_k=e^{\gamma_k t}\frac{q_k}{e_{kk}}\frac{y_k^{e_{kk}/q_k}}{x_k}$$
$$p_k^y=\frac{\partial L}{\partial \dot{y}_k}=G_k=-e^{\gamma_k t}y_k^{e_{kk}/q_k-1} \log x_k$$ These lead us to the primary constraints of the model which are defined as
\be\label{pri}\phi_k:=p_k^x-F_k\approx 0, \;\;\;\psi_k:= p_k^y-G_k\approx 0.\ee
The primary Hamiltonian is therefore
\be \label{priHam}
H_p=H+\sum_{k=1}^m (\lambda_k \phi_k+\mu_k\psi_k), \;\;\;k=1,...,m\ee
As the primary constraints must be satisfied at all times
 it is necessary that their time evolution vanish. This requirement leads us to the second class constraints of the system.
 $$\dot{\phi}_k=\{\phi_k, H\}+\sum_{k=1}^m \mu_k\{\phi_x, \psi_k\}\approx 0$$
 We therefore define the secondary constraints corresponding to the $\phi$-class of primary constraints by
 \be\label{secphi}\phi_{k+m}=\{p_k^x, U\}+\sum_{k=1}^m \mu_k\{\phi_k, \psi_k\}\approx 0, \;\;\;k=1,...,m\ee
 In a similar manner for the $\psi$-class of primary constraints one defines the corresponding secondary constraints as
 \be\label{secpsi}\psi_{k+m}=\{p_k^y, U\}+\sum_{k=1}^m \lambda_k\{\psi_k, \phi_k\}\approx 0, \;\;\;k=1,...,m\ee

Let us now define the matrix of the Poisson brackets between the  all the primary and second class constraints as $C$  which is obviously a skew symmetric
$4m\times 4m$ matrix.
$$C=\left[\begin{array}{cccc}
\{ \phi_i, \phi_j\}_{2m\times 2m} & \{ \phi_i, \psi_j\}_{2m\times 2m}\\
\{ \psi_i, \phi_j\}_{2m\times 2m} & \{ \psi_i, \psi_j\}_{2m\times 2m}\end{array}\right]$$
This turns out to be non-singular and in terms of  its elements the time evolution of any variable $f$ is given by
\be\dot{f}=\{f, H\} -\left(\{f, \phi_1\},\cdots,\{f, \phi_{2m}\}, \{f, \psi_1\}, \cdots \{f, \psi_{2m}\}\right) C^{-1}\left(\begin{array}{ccc} \{\phi_1, H\}\\
\vdots\\
\{\phi_{2m}, H\}\\
\{\psi_1, H\}\\
\vdots\\
\{\psi_{2m}, H\}\end{array}\right).\ee
 We now present the explicit nature of the calculations stated above  by considering the case $m=1$, i.e, for the system (\ref{ktw1a}) and (\ref{ktw1b}). The  JLM it will be recalled is given by
 $J=e^{\gamma t}y^\sigma/x$ where $\gamma=e_{11}d/q$ and $\sigma+1=e_{11}/q$. It follows that the Lagrangian for the system is given by
 $$L=\frac{e^{\gamma t}y^{\sigma+1}}{x(\sigma +1)}\dot{x} -(e^{\gamma t} y^\sigma \log x) \dot{y} -U(x, y, t) $$ where
 $$U(x, y, t)=-e^{\gamma t}y^{\sigma +1}\left[\frac{2p}{\sigma +2} y+2q x -d\log x -\frac{2qb_1}{e_{11}} \right]$$
 The Hamiltonian is then found to be $H=U$ and the primary constraint equations which follow from the usual definition of the conjugate momenta are
 $$\phi_1=p_x-\frac{e^{\gamma t}y^{\sigma +1}}{x(\sigma +1)}\approx 0$$
$$\phi_2=p_y+e^{\gamma t}y^\sigma \log x\approx 0$$
The primary Hamiltonian is therefore given by
$$H_p=U(x, y, t)+\lambda_1\phi_1+\lambda_2\phi_2$$ and from the time evolution of the primary constraints we arrive at the following second-class constraints, namely:
$$\phi_3=-U_x-\lambda_2\left(\frac{2e^{\gamma t}y^\sigma}{x}\right)\approx 0$$
$$\phi_4=-U_y+\lambda_1\left(\frac{2e^{\gamma t}y^\sigma}{x}\right)\approx 0$$ It is now a straightforward matter to calculate the matrix of the Poisson brackets of the primary and second-class constraints and we find that
$$C=\left(\begin{array}{cccc}
0 & -\frac{2e^{\gamma t}y^\sigma}{x} & -\phi_{3x} & -\phi_{4x}\\
\frac{2e^{\gamma t}y^\sigma}{x} & 0 & -\phi_{3y} & -\phi_{4y}\\
\phi_{3x} & \phi_{3y} & 0 & 0\\
\phi_{4x} & \phi_{4y} & 0 & 0\end{array}\right)$$ and its inverse is given by
$$C^{-1}=\frac{1}{\Delta}\left(\begin{array}{cccc}
0 & 0 &\phi_{4y} & -\phi_{3y}\\
0 & 0 & -\phi_{4x} & \phi_{3x}\\
-\phi_{4y} & \phi_{4x}  & 0 & -\frac{2e^{\gamma t}y^\sigma}{x}\\
 \phi_{3y}& -\phi_{3x} &\frac{2e^{\gamma t}y^\sigma}{x} & 0 \end{array}\right)$$
 where $\Delta=\phi_{3x}\phi_{4y}-\phi_{4x}\phi_{3y}$.
 Explicit calculation of the time evolution of the phase space variables $(x, y, p_x, p_y)$  using the Dirac brackets now yields
\be\dot{x}=0,\ee
\be\dot{y}=0,\ee
\be \dot{p}_x=(-1+\Delta)U_x,\ee
\be\dot{p}_y=(-1+\Delta)U_y.\ee
We now turn to certain simplified reductions of the KtW model presented above and deduce explicitly the equations of motion for the phase space variables.
\subsubsection{Lotka-Volterra model with competition}
 This is a model similar to the original Lotka-Volterra model but which incorporates the competition between species which is
  modeled by a term proportional to the product of the populations of the prey and predator \cite{Smale}. The equations are given by
  \be\label{LVCa} \dot{x}=x(a_1-b_1x-c_1y)\ee
  \be\label{LVCb} \dot{y}=y(a_2-b_2y-c_2x)\ee
  where $a_i, b_i, c_i>0\;\;\forall i=1,2$. It is evident by comparison with (3.1) and (3.2) that this model is a special case of the KtW model with an extra term proportional to $ y^2$ in the second equation. The Jacobi Last Multiplier for this system of equations is  given by
  $\mu=e^{\gamma t}x^\alpha y^\beta$ with the exponents being
  $$\alpha=\frac{b_2c_2+c_1c_2-2b_1b_2}{b_1b_2-c_1c_2},$$
  $$\beta=\frac{b_1c_1+c_1c_2-2b_1b_2}{b_1b_2-c_1c_2},$$
  $$\gamma=\frac{a_1(b_1b_2-b_2c_2)+a_2(b_1b_2-b_1c_1)}{b_1b_2-c_1c_2}.$$ It turns out that a singular Lagrangian for the above system is given by
  $$L=-e^{\gamma t}\frac{x^\alpha y^{\beta+1}}{\beta+1}\dot{x}+e^{\gamma t}\frac{x^{\alpha+1} y^{\beta}}{\alpha+1}\dot{y}-V(x,y, t),$$
  where
  $$V(x, y, t)=e^{\gamma t}x^{\alpha+1}y^{\beta+1}\left[\frac{2(a_2-b_2y)}{\alpha+1}+\frac{\gamma}{(\alpha+1)(\beta+1)}-\frac{2c_2x}{\alpha+2}\right],$$  or alternately
  $$=e^{\gamma t}x^{\alpha+1}y^{\beta+1}\left[-\frac{2(a_1-b_1x)}{\beta+1}-\frac{\gamma}{(\alpha+1)(\beta+1)}+\frac{2c_1y}{\beta+2}\right].$$
  A characteristic feature of singular Lagrangians is that the Hamiltonian is given by
  $$H=p_x\dot{x}+p_y\dot{y}-L=V(x, y, t)$$ and is therefore independent of the velocities.
  The primary constraints are therefore
  $$\phi_1=p_x-F(x, y, t)\approx 0,$$
  $$\phi_2=p_y-G(x, y, t)\approx 0,$$ where
  $$F(x, y, t)=-e^{\gamma t}\frac{x^\alpha y^{\beta+1}}{\beta+1},$$
  $$G(x, y, t)=e^{\gamma t}\frac{x^{\alpha+1} y^{\beta}}{\alpha+1},$$ respectively. Hence the primary Hamiltonian is
  $$H_p=V+\lambda_1 \phi_1+\lambda_2 \phi_2.$$
  On the other hand the second class constraints are given by
  $$\phi_3=\dot{\phi}_1=\{\phi_1, H_p\}=\{p_x, V(x,y, t)\}+\lambda_2 \{\phi_1, \phi_2\}\approx 0,$$
  $$\phi_4=\dot{\phi}_2=\{\phi_2, H_p\}=\{p_y, V(x,y, t)\}+\lambda_1\{\phi_2, \phi_1\}\approx 0,$$
that is
$$\phi_3=-V_x+\lambda_2(2e^{\gamma t}x^\alpha y^\beta)\approx 0,$$
$$\phi_4=-V_y-\lambda_1(2e^{\gamma t}x^\alpha y^\beta)\approx 0.$$
The matrix of the Poisson brackets between the constraints is given by
$$C=\left(\begin{array}{cccc}
0 & 2e^{\gamma t}x^\alpha y^\beta & V_{xx}-\alpha V_x/x & V_{xy}+\alpha V_x/x\\
-2e^{\gamma t}x^\alpha y^\beta & 0 & V_{xy}+\beta V_y/y & V_{yy}-\beta V_y/y\\
-V_{xx}+\alpha V_x/x & -V_{xy}-\beta V_y/y & 0 & 0\\
-V_{xy}-\alpha V_x/x & -V_{yy}+\beta V_y/y & 0 &0\end{array}\right).$$
This is a non-singular matrix and its inverse is
$$C^{-1}=\frac{1}{\Delta}\left(\begin{array}{cccc}
 0 & 0 &-V_{yy}+\beta V_y/y & V_{xy}+\beta V_y/y\\
 0 & 0 & V_{xy}+\alpha V_x/x & -V_{xx}+\alpha V_x/x\\
 V_{yy}-\beta V_y/y & -V_{xy}-\alpha V_x/x & 0 & 2e^{\gamma t}x^\alpha y^\beta\\
  -V_{xy}-\beta V_y/y   &  V_{xx}-\alpha V_x/x & -2e^{\gamma t}x^\alpha y^\beta & 0\end{array}\right),$$
  where $\Delta=(V_{xx}V_{yy}-V_{xy}^2) -\alpha V_x(V_{xy}+V_{yy})/x -\beta V_y(V_{xy}+V_{xx})/y$. The time evolution of a dynamical variable $f$, in terms of the Dirac brackets, is
  $$\dot{f}=\{f, H\}_{DB}=\{f, H\} -\{f, \phi_i\}[C^{-1}]^{ij} \{\phi_j, H\},$$ and upon using the entries of the matrix $C^{-1}$ as given above we find that
  $$\dot{x}=0,$$
  $$\dot{y}=0,$$
 $$ \dot{p}_x=-V_x-\frac{1}{\Delta}\left[(V_{xx}-\frac{\alpha V_x}{x})\{(-V_{yy}+\frac{\beta V_y}{y})V_x+(V_{xy}+\frac{\alpha V_x}{x})V_y\}\right.$$
$$ \left.+(V_{xy}-\frac{\alpha V_y}{x})\{(-V_{xy}+\frac{\beta V_y}{y})V_x+(-V_{xx}+\frac{\alpha V_x}{x})V_y\} \right]$$
 $$ \dot{p}_y=-V_y-\frac{1}{\Delta}\left[(V_{xy}-\frac{\beta V_x}{y})\{(-V_{yy}+\frac{\beta V_y}{y})V_x+(V_{xy}+\frac{\alpha V_x}{x})V_y\}\right.$$
$$\left. +(V_{yy}-\frac{\beta V_y}{y})\{(V_{xy}+\frac{\beta V_y}{y})V_x+(-V_{xx}+\frac{\alpha V_x}{x})V_y\} \right]$$
\subsubsection{Lotka-Volterra model without competition}

In the absence of competition, the coefficients ($b_1$, $b_2$) in the Lotka-Volterra model with competition vanish and a simplified version of the predator-prey systems becomes
\begin{eqnarray}
\dot{x}=ax-bxy\label{a}\\
\dot{y}=-cy+dxy\label{b}\end{eqnarray}
 where $x$ and $y$ refer to two species which live in a limited area with individual of the species $y$(predator) feed only on the species $x$ (prey). The parameters $a, b, c$ and $d$ are assumed to be positive. The overdots refer to derivatives with respect to time with $\dot{x}$ representing the growth rate of the prey population and $\dot{y}$ being the growth rate of the predator population. This model can also be visualized as the special case of the KtW model with the absence of the term proportional to $x^2$ in the $\dot{x}$ equation.\\
 \\
 One can recast the system (\ref{a}) and (\ref{b}) in the form of Euler-Lagrange equation with the Lagrangian
 \be\label{Lag1} L=\left(-\frac{\log x}{y}\dot{x}+\frac{\log y}{x}\dot{y}\right)-c\log x +a\log y +dx -by,\ee and the corresponding  momenta are
 \be\label{mom}p_x=\frac{\partial L}{\partial \dot{x}}=-\frac{\log{y}}{x}, \;\;\;p_y=\frac{\partial L}{\partial \dot{y}}=\frac{\log x}{y}.\ee
 The Hamiltonian is therefore  given by
 \be\label{LVHam} H=-2c\log x-2a\log y +2dx +2by.\ee
 The Hamiltonian is clearly independent of the momentum (or velocity) and is therefore of singular character. In order to  study the dynamics in the phase space  $(x, y, p_x, p_y)$ we treat the momenta as given in (\ref{mom}) as the primary constraints and express these in the form
 \be\label{priconst}\phi_1=p_x+\frac{\log y}{x}\approx 0,\;\;\;\phi_2=p_y-\frac{\log x}{y}\approx 0\ee
 The primary Hamiltonian is then defined by
 \be\label{priHam} H_p=H+\lambda_1\phi_1+\lambda_2\phi_2,\ee where $\lambda_1$ and $\lambda_2$ are Lagrange multipliers. In order to ensure that the constraints hold at all times it is necessary that their time evolutions with respect to the Hamiltonian $H_p$ vanish. In other words we require that
 $$\dot{\phi}_1=\{ \phi_1, H\}=\frac{2c}{x}-2d +2\frac{\lambda_2}{xy}\approx 0,$$
 $$\dot{\phi}_2=\{ \phi_2, H\}=\frac{2a}{y}-2b -2\frac{\lambda_1}{xy}\approx 0.$$
 These furnish us with two more constraints which are referred to as the secondary constraints or  constraints of the second class namely
 \be\label{seccons}\phi_3=\frac{2c}{x}-2d +2\frac{\lambda_2}{xy}\approx 0, \;\;\;\phi_4= \frac{2a}{y}-2b -2\frac{\lambda_1}{xy}\approx 0.\ee
 Thus we have in all four constraints. Let $C$ denote that matrix formed by the Poisson brackets between $\phi_i$ and $\phi_j$, i.e,  $C_{ij}=\{\phi_i, \phi_j\}$. The $\approx$ sign here means we can only substitute the value of $\lambda_i$ from the secondary constraint equations after working out the respective Poisson brackets.  With this in mind it is found that
\be C=\frac{2}{xy}\left(\begin{array}{ccccc}
 0 & 1& dy & -(a-by)\\
 -1 & 0 & -(c-dx) & bx\\
 -dy & (c-dx) & 0 & 0\\
 (a-by) & -bx & 0 & 0\end{array}\right).\ee  It may be verified that $C$ is non-singular and its inverse is given by
 \be C^{-1}=\frac{2}{xy\xi}\left(\begin{array}{ccccc}
0 & 0& -bx & -(c-dx)\\
 0 & 0 & -(a-by) & -dy\\
 bx & (a-by) & 0 & 1\\
 (c-dx) & dy & -1 & 0\end{array}\right),\ee
 where $\det C=\xi=adx +bc y-ac$.
 The time evolution of any dynamical quantity in terms of the Dirac bracket is defined as
 \be\label{DB} \dot{f}=\{ f, H\}_{DB}=\{f, H\}-\{f, \phi_i\}[C^{-1}]^{ij}\{\phi_j, H\}\ee
 Using this definition we find in the present case the following equations of motion.
 \begin{eqnarray}
 \dot{x}=0, \;\;\;
 \dot{p}_x=\frac{2(c-dx)}{x}\left(1-\frac{4}{x^2y^2}\right),\\
 \dot{y}=0,\;\;\;
 \dot{p}_y=\frac{2(a-by)}{y}\left(1-\frac{4}{x^2y^2}\right).\end{eqnarray}

\subsubsection{Kermack-McKendreck model}
A simplified version of  this model which is often cited  in the literature is
 given by the following system of differential equations \cite{Ruan,GCG1}:
 \begin{eqnarray}
 \dot{x}=-k_1 xy\\
 \dot{y}=k_1 xy -k_2 y\end{eqnarray}
 with $k_1$ and $k_2$ being positive constants. This model can also be derived from the KtW model by suppressing some coefficients. The system may be derived from the Euler-lagrange equations with the Lagrangian
 \be\label{KMacLag}L=\frac{1}{2}\left(\frac{\log y}{x}\dot{x}-\frac{\log x}{y}\dot{y}\right)+k_1(x+y)-k_2\log x\ee
 The conjugate momenta are obtained from
 $$p_x=\frac{\partial L}{\partial \dot{x}}=\frac{\log y}{2x},$$
 $$ p_y=\frac{\partial L}{\partial \dot{y}}=-\frac{\log x}{2y},$$ and are obviously velocity independent as the Lagrangian is singular in character.
 This necessitates that we defined the primary constraints by
 \be\label{Pcons}\phi_1=p_x-\frac{\log y}{2x}\approx 0, \;\;\;\phi_2=p_y+\frac{\log x}{2y}\approx 0.\ee
 The Hamiltonian therefore has the appearance
 \be\label{KMacHam}H=p_x\dot{x}+p_y\dot{y}-L=-k_1(x+y)+k_2\log x.\ee It is easy to verify that $H$ is a constant of motion.
 We define the primary Hamiltonian by
 \be\label{PKMacHam} H_p=H+\lambda_1\phi_1+\lambda_2\phi_2.\ee
 In order to ensure that the primary constraints hold at all times it is necessary that
 $\dot{\phi}_i=\{\phi_i, H_p\}$ $i=1,2$ vanish. This in turn leads us to the secondary constraints, namely
 $$\dot{\phi}_1=\{\phi_1, H_p\}=\{\phi_1, H\}+\lambda_2\{\phi_1, \phi_2\}$$ whence we have
 \be\label{Scons1}\phi_3=k_1-\frac{k_2}{x}-\frac{\lambda_2}{xy}\approx 0.\ee
 Similarly considering the vanishing of the time evolution of $G_2$ we find that
 \be\label{Scons2}\phi_4=k_1+\frac{\lambda_1}{xy}\approx 0.\ee

 The matrix of the Poisson brackets of the primary and secondary constraints is then given by
 \be\label{emmm}
 C=\left(\begin{array}{cccccc}
 0 & -\frac{1}{xy} & -\frac{k_2}{x} & -\frac{k_1}{x}\\
 \frac{1}{xy} & 0 &-\frac{k_1}{y}+\frac{k_2}{xy} & -\frac{k_1}{y}\\
  \frac{k_1}{x} & \frac{k_1}{y}-\frac{k_2}{xy} & 0 & 0\\
  \frac{k_1}{x} & \frac{k_1}{y} & 0 & 0\end{array}\right)\ee
  This is a non-singular matrix and its inverse is given by
  \be\label{invem}C^{-1}=\frac{x}{k_1k_2}\left(\begin{array}{cccc}
  0 & 0 & k_1x & -k_1x+k_2\\
  0 & 0 & -k_1y & k_1y\\
  -k_1x & k_1y & 0 & -1\\
  k_1x-k_2 & -k_1y &1 & 0\end{array}\right)\ee

 If we go back to the definition of the Dirac bracket it will be realized that
 the equations of motion of the phase space variables are now obtained from
 $$\dot{f}=\{f, H\}_{DB}=\{f, H\}-\frac{1}{k_1k_2}(\{f, \phi_1\}, \{f, \phi_2\}, \{f, \phi_3\}, \{f, \phi_4\})\left(\begin{array}{cc}
 0\\0\\-k_1^2x(x-y)+k_1k_2x\\(k_1x-k_2)^2-k_1^2xy\end{array}\right)$$
 It may be verified that this implies
 $$\dot{x}=0$$
 $$\dot{y}=0$$
$$\dot{p}_x=(k_1k_2-1)\left(-k_1+\frac{k_2}{x}\right)$$
$$\dot{p}_y=k_1-k_1^2k_2$$

\subsection{Gierer-Meinhardt Model}
This is a reaction-diffusion model which deals with the formation of patterns \cite{GM,GPP, Turing}. A simplified version of this model neglecting the effects of diffusion is given by
\begin{eqnarray}
\dot{x}=x^2y-a x,\\
\dot{y}=b-x^2y-y.
\end{eqnarray}
In order to express this system in the form of Euler-Lagrangian equations, it is found that unless the parameter $b=0$, we cannot find a Lagrangian.
Simple calculations show that with $b=0$ one can derive a Jacobi Last Multiplier for this system which is given by
$$\mu=\frac{e^{-at}}{x^2y}$$ and the Lagrangian (singular) is of the form
$$L=e^{-at}\left[\frac{\log y}{x^2}\dot{x}+\frac{1}{xy}\dot{y}-\left(2(x-\frac{1}{x})+2y-a\frac{\log y}{x}\right)\right]$$
The Hamiltonian is then given by
$$H=e^{-at}\left[2(x-\frac{1}{x})+2y-a\frac{\log y}{x}\right]$$
The primary constraints are therefore given by
$$\phi_1=p_x-e^{-at}\frac{\log y}{x^2}\approx 0$$
$$\phi_2=p_y-e^{-at}\frac{1}{xy}\approx 0$$
Consequently we define the primary Hamiltonian as
$$H_p=H+\lambda_1\phi_1+\lambda_2\phi_2.$$
By demanding  that the time evolution of the primary constraints vanish we are led to the following secondary constraints, as already explained earlier, namely
$$\phi_3=e^{-at}\left(2+\frac{2}{x^2}+\frac{a}{x^2}\log y +\frac{2\lambda_2}{x^2y}\right)\approx 0,$$
$$\phi_4=e^{-at}\left(\frac{a}{xy}-2+\frac{2\lambda_1}{x^2y}\right)\approx 0.$$
The matrix of the Poisson brackets of the primary and secondary constraints is given by
$$C=\frac{2e^{-at}}{x^2y}\left(\begin{array}{cccccc}
0 & -1 & -2xy & 2xy-\frac{a}{2}\\
1 & 0 & \epsilon & x^2\\
2xy & -\epsilon & 0 & 0\\
-2xy+\frac{a}{2} & -x^2 &0 & 0\end{array}\right),$$ where
$$\epsilon =-(x^2+1+\frac{a}{2} +\frac{a}{2}\log y).$$
 Straightforward calculations give the inverse of the matrix $M$ to be
 $$C^{-1}=    \frac{e^{at} x^2 y}{2\delta }\left(\begin{array}{cccc}
 0 & 0 & -x^2 & \epsilon\\
  0 & 0 & -2xy-\frac{a}{2} & 2xy\\
  x^2 & -2xy+\frac{a}{2} & 0 -1\\
  -\epsilon & -2xy & 1 & 0\end{array}\right),$$
  where
  $$\delta=2x^3y-\epsilon (2xy-\frac{a}{2})$$
 The calculation of the Dirac brackets and the equations of motion of the phase space variables follow in the usual manner as in the earlier examples.
\section{Conclusion}

In the diverse field of microbial systems where the mutation rates of different species are very high and also in the field of quantitative biology like prey-predator system, host-parasite model etc. where the prey predator relationships becomes much more complicated with respect to the time parameter exact analytical  results are often rare. In this paper we have attempted to provide a Lagrange/Hamiltonian description of an important stochastic model, popularly referred to as the "Kill the Winner" model. It is evident that such a description is facilitated by the existence of a Jacobi Last Multiplier which is not very well known outside the community of mathematicians working on systems of ordinary differential equations. Its existence leads quite naturally to a Lagrange description of the model equations, which in the present case turns out to be of a singular character.  As a result the corresponding Hamiltonian description of the phase space variables requires the introduction of Dirac brackets. We have presented explicit results for a number of models such a predator-prey systems with and without competition, pattern formation equations  and of course the KtW model which forms the cornerstone of the article.

\section*{Acknowledgement}
PG is extremely grateful to Pepin Carinena and Ogul Esen for enlighting discussions and valuable comments.


\begin{thebibliography}{99}
\bibitem{KTW} Chi Xue and Nigel Goldenfeld, Phy. Rev. Lett. 119 (2017), 268101.
\bibitem{TF} S.L. Trubatch and A. Franco, {\em Canonical Procedures for Population Dynamics}, J. Theor. Biol. 48 (1974), 299-324.
    \bibitem{Go} A. Goriely, {\em Integrability and nonintegrability of dynamical systems}.
Advanced Series in Nonlinear Dynamics, 19. World Scientific
Publishing Co., Inc., River Edge, NJ, 2001. xviii+415 pp.
\bibitem{Paine} G.H. Paine, {\em The development of Lagrangians for biological models},
Bull. Math. Biol. 44 (1982) 749-760.
\bibitem{NT}M.C.Nucci and K M Tamizhmani {\em Lagrangians for
biological systems}, J. Nonlinear Math. Phys (19) (3) 1250021 (2012).
\bibitem{Dirac} P. A. M. Dirac, {\em Generalized Hamiltonian Dynamics}, Canadian J. Math 2 (1950) 129-148.
\bibitem{GN} M. J. Gotay, J. M. Nester, and G. Hinds, {em Presymplectic mani-folds and the Dirac-Bergmann theory of constraints},	J. Math. Phys 19 (1978) 2388-2399.
\bibitem{esenguha} O. Esen and P. Guha, {\em On time-dependent Hamiltonian realizations of planar and nonplanar systems},
	J. Geom. Phys. 127 ( 2018 ) 32-45.
\bibitem{deLeon1} M. de Leon and D. Martin de Diego, {\em A constraint algorithm for singular Lagrangians subjected to nonholonomic constraints},
J. Math. Phys. 38 (1997) 3055-3062.
\bibitem{deLeon2} M. de Leon and C. Sardon, {\em Cosymplectic and contact structures for time-dependent and dissipative Hamiltonian systems},
	J. Phys A: Mathematical and Theoretical 50 (2017) 255205.

\bibitem{Saunders}  D. J. Saunders, {\em The Geometry of Jet Bundles}, LMS vol \textbf{142},
Cambridge University Press, Cambridge, 1989.
\bibitem{Ibort}   A. Ibort, M. de Leon, J. C. Marrero, and D. Martin de Diego, {\em Dirac brackets in constrained dynamics},
	Fortschritte der Physik 47 (1999) 459-492.		
\bibitem{Car1} J.F. Cari\~nena and J. Fernandez-Nu\~nez, {\em Geometric Theory of Time-Dependent
Singular Lagrangians}, Fortschritte der Physik \textbf{41}, Issue 6 (1993) 517–552.
\bibitem{Car2} J.F. Cari\~nena, J. Fernandez-Nu\~nez and M. Ra\~nada, {\em Singular Lagrangians
affine in velocities}, J. Phys. A: Math. Gen. \textbf{36} (2003) 3789.
\bibitem{Crampin} M.Crampin, {\em Jet Bundle Techniques in Analytical Mechanics},
(Gruppo Nazionale di Fisica Matematica) vol \textbf{47} (1995).
\bibitem{Whi} E.T. Whittaker, {\em A Treatise on the Analytical Dynamics of Particles
and Rigid Bodies}. Cambridge University Press, Cambridge, 1988.
\bibitem{Leach}M.C.Nucci and P.G.L.Leach, {\em Jacobi's last
multiplier and  Lagrangians for multidimensional
systems.}\textit{arXiv:0709.3231v1}.

\bibitem{Nucci} M.C.Nucci, {\em Jacobi last multiplier and Lie symmetries: a novel application of an old
relationship}, J. Nonlinear Math. Phys. 12 (2005) 284–304.

\bibitem{CGK} A. Ghose Choudhury, P. Guha and B. Khanra, {\em On the Jacobi last multiplier, integrating
factors and the Lagrangian formulation of differential equations
of the Painlev\'e-Gambier classification}, J. Math. Anal. Appl.
360  (2009),  no. 2, 651--664.
\bibitem{CG} A. Ghose Choudhury and P. Guha, {\em Application of Jacobi's last multiplier for
construction of Hamiltonians of certain biological systems}, Cent. Eur. J. Phys. • 10(2) • 2012 • 398-404.
\bibitem{Ruan} S. Ruan, {\em Diffusion-driven instability in the Gierer-Meinhardt model of morphogenesis},
Natural Resourse Modeling 11, no. 2, summer 1998.
\bibitem{GCG1} P. Guha and A. Ghose Choudhury, {\em Singular Lagrangian, Hamiltonization and
Jacobi Last Multiplier for Certain Biological Systems}, Eur. Phys. J. Special Topics 2013, Volume 222, Issue 3-4, 615-624.
\bibitem{Smale} S Smale, {\em On the differential equations of species in competition}, J. Math. Biol (2016), 3 (2) 5-7.

\bibitem{GM} A. Gierer and H. Meinhardt, {\em A theory of biological pattern formation}.
 Kybernetik 12 (1972) 30-39.
\bibitem{GPP} M.I. Granero-Porati and A. Porati, {\em Temporal organization in a morphogenesis field},
J. Math. Biol 20 (1984) 153-157.


\bibitem{Turing} A.M. Turing, {\em The chemical basis of morphogenesis}.
Phil. Trans. R. Soc. Lond. B237, 37-72 (1952).




\end{thebibliography}
 \end{document}